\documentclass[11pt,a4paper]{article}
\pdfoutput=1

\usepackage{jheppub}
\usepackage{amsfonts}
\usepackage{amsmath}
\usepackage{amssymb}
\usepackage{bm}
\usepackage{dcolumn}
\usepackage{epsfig}
\usepackage{graphicx}
\usepackage{graphics}
\usepackage[latin1]{inputenc}
\usepackage{latexsym}
\usepackage{rotating}
\usepackage{hyperref}
\usepackage{dsfont}

\usepackage{color}
\usepackage[active]{srcltx}
\usepackage{amsmath,amsfonts,amssymb,amsthm,amstext,amscd,eucal,srcltx}
\usepackage{epsfig,graphicx,bm}
\usepackage{epstopdf, epsf}
\usepackage{dcolumn}
\usepackage{hyperref}

\newcommand{\bea}{\begin{eqnarray}}
\newcommand{\eea}{\end{eqnarray}}

\usepackage{amssymb} \usepackage{amsmath} \usepackage{graphicx}
\usepackage{epsfig,latexsym}

\def\bef{\begin{figure}}
\def\eef{\end{figure}}

\newcommand{\be}[1]{\begin{equation}\label{#1}}
\newcommand{\beq}{\begin{equation}}
\newcommand{\ee}{\end{equation}}
\newcommand{\beqn}[1]{\begin{eqnarray}\label{#1}}
\newcommand{\eeqn}{\end{eqnarray}}
\newcommand{\bd}{\begin{displaymath}}
\newcommand{\ed}{\end{displaymath}}

\def\lsim{\raise0.3ex\hbox{$\;<$\kern-0.75em\raise-1.1ex
e\hbox{$\sim\;$}}}
\def\gsim{\raise0.3ex\hbox{$\;>$\kern-0.75em\raise-1.1ex
\hbox{$\sim\;$}}}
\def\simlt{\mathrel{\lower2.5pt\vbox{\lineskip=0pt\baselineskip=0pt
           \hbox{$<$}\hbox{$\sim$}}}}
\def\simgt{\mathrel{\lower2.5pt\vbox{\lineskip=0pt\baselineskip=0pt
           \hbox{$>$}\hbox{$\sim$}}}}
\def\unity{{\hbox{1\kern-.8mm l}}}

\newcommand{\vect}[1]{\mbox{\boldmath$#1$}}

\def\lsim{\mathrel{\mathop  {\hbox{\lower0.5ex\hbox{$\sim$}
\kern-0.8em\lower-0.7ex\hbox{$<$}}}}}
\def\gsim{\mathrel{\mathop  {\hbox{\lower0.5ex\hbox{$\sim$}
\kern-0.8em\lower-0.7ex\hbox{$>$}}}}}

\title{Self-dual formulation of gravity in topological M-theory}

\vspace{0.1cm}
\author[a]{Andrea Addazi}

\author[a]{and Antonino Marcian\`o}
\affiliation[a]{Center for Field Theory and Particle Physics \& Department of Physics, Fudan University, 200433 Shanghai, China}

\begin{document}

\vspace*{3mm}

\abstract{
\noindent
Inspired by the low wave-length limit of topological M-theory, which re-constructs the theory of $3+1$D gravity in the self-dual variables' formulation, and by the realization that in Loop Quantum Gravity the holonomy of a flat connection can be non-trivial if and only if a non-trivial (space-like) line defect is localized inside the loop, we argue that non-trivial gravitational holonomies can be put in correspondence with space-like M-branes. This suggests the existence of a new duality, which we call $H$ duality, interconnecting topological M-theory with Loop Quantum Gravity. We spell some arguments to show that fundamental S-strings are serious candidates to be considered in order to instantiate this correspondence to classes of LQG states. In particular, we consider the case of the holonomy flowers in LQG, and show that for this type of states the action of the Hamiltonian constraint, from the M-theory side, corresponds to a linear combination of appearance and disappearance of a SNS1- strings. Consequently, these processes can be reinterpreted, respectively, as enucleations or decays into open or closed strings. 
}

\maketitle

\section{Introduction}

\noindent 
M-theory and Loop Quantum Gravity (LQG) are usually accounted as the two most prominent candidates to solve the problem of non-renormalizability of quantum gravity for energies higher than the Planck scale. It is commonly retained that these theories cannot be compatible with one another. Without any direct experimental data on the string/Planck scale, both the theories can only follow mathematical self-consistency. Indeed, both the frameworks are still concerned with several technical and conceptual problems, while use known tools of quantum field theory that physicists trust by virtue of the experimental successes of the Standard Model of particles and interactions. 

Nonetheless the history of string theory, in which many different models were connected one another by dualities, suggests that string theory and LQG might actually be unified despite the profound differences they have. 
\begin{figure}[t!!]
\begin{center}
\vspace{1cm}
\includegraphics[width=12cm,height=8cm,angle=0]{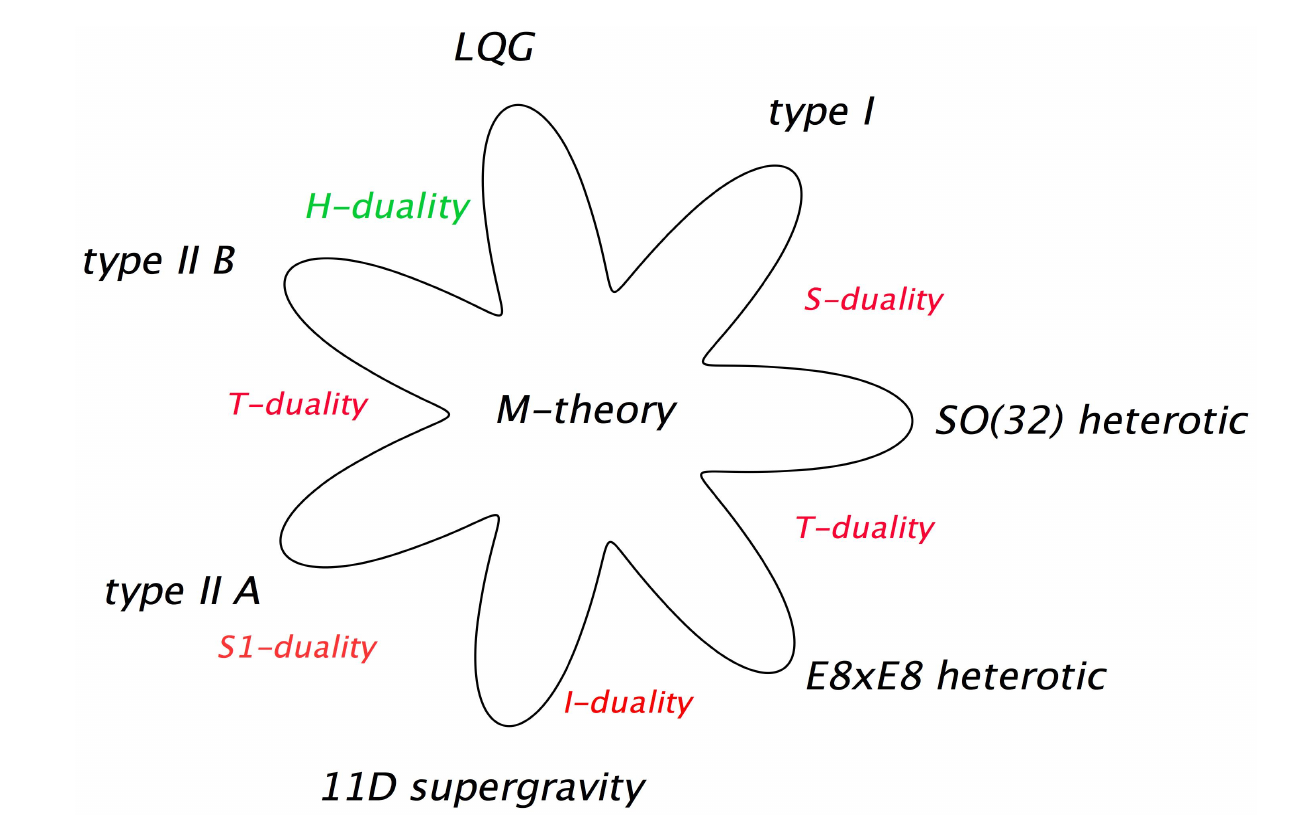}
\caption{We show the seven epiphenomena of non-perturbative M-theory. In particular, we conjecture that LQG is recovered from the duality among holonomies in the Hilbert space of LQG and 
space-like fundamental strings in IIB-string theory, related to SM2-branes by a double reduction. 
We dub this new duality H-duality (written in green and located between IIB-type and LQG). Other duality transformations hitherto discovered are also represented in red. } \label{fig:1}
\end{center}
\end{figure}
This urges posing the question: {\it can M-theory and LQG describe aspects of the same theory from two different points of view?} Reasons to prefer M-theory to LQG are usually individuated in the unification of all particles and interactions, and in the convergence --- at the low energy perturbative limit --- to General Relativity (GR). On the other hand, LQG has the remarkable advantage to be background independent and a fully non-perturbative theory\footnote{Unification models were also proposed within the framework of LQG. See {\it e.g.} \cite{BilsonThompson:2006yc,Alexander:2011jf,Alexander:2012ge}.}. It seems therefore to be very challenging to reconcile theories that are so much different one another: M-theory requires supersymmetry and higher dimensions, while in LQG the Ashtekar variables were originally defined only in 4 dimensions and without the need of supersymmetry. Nonetheless, supersymmetric and higher dimensional reformulations of LQG have been also considered in the literature, even if their dynamics was not deepened in detail as for the standard formulation of LQG.

Recently, a promising unification approach between string theory and LQG was proposed in Refs.~\cite{Turiaci:2017zwd,Mertens:2017mtv}, within the context of holographic $AdS_{2}/CFT$ and $AdS_{3}/CFT$ models. In these works, the gravitational scattering matrix of a particle crossing the horizon, treated in the optical/eikonal approximation, is related to the Schwarzian quantum mechanical model in the large N approximation. This is in turn related to considering $6j$-symbol of $SU(1,1)$. At the present stage, it is still unknown how to generalize such correspondences among string theories and LQG in higher dimensions.


\vspace{0.5cm}
In this paper, we focus on the possible correspondence principle between LQG and M-theory. This is a conjecture sustained by two insightful facts. 
First, topological M-theory on 7-dimensional $G_{2}$ manifold has a low wave-length description in terms of 3-form $\Phi$ \cite{Dijkgraaf:2004te} --- this is a result based on seminal papers by Hitchin \cite{Hitchin:2000jd,Hitchin:2001rw}. Then, thanks to a foliation of $G_{2}$ into four dimensional $M$ sub-manifold, one can demonstrate that $\Phi$ corresponds to the self-dual formulation of gravity \cite{Dijkgraaf:2004te}. This may suggest the following conjecture: the quantization procedure deployed in LQG captures aspects of the non-perturbative M-branes dynamics. In particular, the Hamiltonian formulation of LQG may provide a description the space-like M-branes, dubbed SM-branes.
Second, it was noticed that the holonomy of a flat connection can be non-trivially recognized to be different than unity, if and only if a non-trivial (space-like) line defect is localized inside the loop. This is true while considering generic holonomies \cite{Bianchi:2009tj}, and it corresponds to a sort of Aharanov-Bohm effect of the self-dual gravitational field. We are tempted to suggest that this defect can be identified with space-like strings. These objects can be indeed understood as one dimensional S-branes, while these latter completely break all supersymmetric generators, eventually allowing for a non-supersymmetric space-time foam approach. \\

We will argue how spin-network states can be put in correspondence with SM-branes in the Hamiltonian formulation, while spinfoam boundaries can be individuated as M-branes in the covariant formulation. In particular, since (S)M-branes intersections and interactions are described by $\Phi$ $3$-form at the semiclassical level, their intersections/interactions can be labelled as fundamental representations of the internal SU(2) gravitational group. Such a formulation allows to set up a correspondence between a class of SM-branes and a class of spinfoam boundary states. This entails introducing a new duality: the S-brane system is dual to a holonomy state in the LQG Hilbert space. For these reasons, we refer to it as {\it Hilbert}-duality, or also $H$-duality. In particular, we individuate a sub-set of holonomy trees in correspondence with a class of non-intersecting SM-branes. These trees can be shrunk with homotopy transformations of trees into loop holonomy flowers. Flowers can be put in correspondence with a set space-like strings encircled by petals. In this formulation the scalar constraint gets a new intriguing interpretation. It amounts to the quantum superpositions of flowers in which petals have been added or removed. This can be interpreted either as an appearance or as a disappearance of S-branes, i.e. either as a S-brane decay or as a nucleation process. In other words, the scalar constraint can be reinterpreted as describing the time evolution of the S-brane system. Since S-branes are unstable, the first process can be reinterpreted as the decay into open or closed strings within the background, while the nucleation process is expected to happen in the non-perturbative regime.

\section{Topological M-theory} 

\noindent 
We start considering the formulation of topological M-theory in 7-dimensions as discussed in Ref.~\cite{Dijkgraaf:2004te}, which stands a preliminary study in developing a unified picture of all \emph{D}-dimensional form theories of gravity. Within this framework, the \emph{7}-dimensional topological M-theory generates the topological string theories, as well as the gravitational form theories in \emph{3} and \emph{4} dimensions follow as reductions of the 7-dimensional form theory near associative and coassociative cycles. 

Form theories of gravity naturally lead to calibrated geometries\footnote{Calibrated geometries are \emph{D}-dimensional Riemannian manifolds that are equipped with a calibration $\Phi$, a differential p-form --- for some $0 \leq p \leq D$ --- which is closed, namely $d\Phi=0$, and reduces on a \emph{p}-dimensional subspace to a volume form, {\it i.e.} for any $x\in M$ and for any oriented dimensional subspace $\xi\in T_x M$, the restriction of the calibration satisfies $\Phi|_{\xi}=\lambda {\rm vol}\xi$, with $\lambda\leq1$.  }, a natural setting for the definition of supersymmetric cycles where branes can be wrapped. Form theories can be then understood in terms of counting the BPS states of wrapped branes of superstrings. As a consequence of this picture, an attractor mechanism relating the charges of the black hole (the homology class of the cycle they wrap on) to specific moduli of the internal theory (determining the metric of the internal manifold) can be recovered in the superstring: this stands as a special case of the general idea of obtaining metrics from gravity forms.

As clarified in Ref.~\cite{Dijkgraaf:2004te}, topological strings can be accounted for on Calabi-Yau \emph{3}-folds, {\it i.e.} topological string computations can be embedded into the superstring. Dualities of the superstring, with a natural geometric interpretation in M-theory, can be related to some dualities in topological theories, with a similar geometric explanation in topological M-theory. Thus a natural definition of topological M-theory is that it should be a theory with one extra dimension relative to the topological string, which brings to a \emph{7}-dimensional theory. This means that a M-theory on $M \times S^1$ is equivalent to a topological theory of strings on $M$, where $M$ is the Calabi-Yau manifold. 

When the size $S^1$ is no more constant, one is finally led to a more general \emph{7}-dimensional manifold. Nonetheless, the only manifolds that preserve supersymmetry and are purely geometric are the class of $G_2$ holonomy spaces. Furthermore, M-theory on a $G_2$ holonomy manifold X provided with a U(1) action is equivalent to the Type IIA superstring. A precise definition of topological M-theory on X has been provided in Ref.~\cite{Dijkgraaf:2004te}, as ``{\it the theory equivalent to A model topological strings on X/U(1), with Lagrangian D-branes inserted where the circle fibration degenerates''}. Worldsheets of the A model can end on the Lagrangian branes, while whenever they are lifted up to the full geometry of X, they correspond to closed \emph{3}-cycles to be identified with membrane worldvolumes. Furthermore, string worldsheets which wrap holomorphic cycles of the Calabi-Yau lift to membranes wrapping associative \emph{3}-cycles of the $G_2$ holonomy manifold. As a consequence, the topological M-theory should be classically equivalent to a theory of $G_2$ holonomy metrics, with quantum corrections provided by membranes wrapping associative \emph{3}-cycles.\\

In the next sections, our focus will be then on the s-branes that can be derived by the \emph{4}-dimensional reduction of the \emph{7}-dimensional M-theory of Ref.~\cite{Dijkgraaf:2004te}. In the next subsection, we will introduce $G_2$ manifolds.

\subsection{$G_2$ manifolds}
\noindent 
A $G_2$ manifold is a \emph{7}-dimensional Riemannian manifold \cite{G2,G2a,G2b,G2c}, whose holonomy group is contained in $G_2$ --- see {\it e.g.} Ref.~\cite{Joyce}. This latter is an exceptional simple Lie group that can be described as the automorphism group of the octonions. Equivalently, the $G_2$ group can be introduced as the proper subgroup of the special orthogonal group SO(7) that preserves an \emph{8}-dimensional spinor, or as the subgroup of the general linear group GL(7) that preserves the non-degenerate \emph{3}-form $\Phi$. In general, a manifold is endowed with a $G_2$-structure if each of its tangent spaces can be identified smoothly with the imaginary octonions $\mathfrak{Im}(\mathbb{O})\simeq \mathbb{R}^7$. This is reminiscent of an almost Hermitian manifold, in which each of its tangent spaces can be identified in a smoothly varying way with $\mathbb{C}^m$ (together with its Euclidean inner product). The non-degenerate \emph{3}-form $\Phi$ is an associative form, with Hodge dual $G=\star \Phi$ representing a parallel, coassociative, \emph{4}-form. These two forms represent the calibrations of the manifold \cite{HL}, and also define special classes of \emph{3}-dimensional and \emph{4}-dimensional submanifolds. 

Notice also that:

\begin{itemize}

\item

A manifold can admit a $G_2$-structure, if the following conditions, which are necessary and sufficient, are satisfied: it is orientable and spin-structured. These are equivalent to the vanishing of its first two Stiefel-Whitney classes, and allow to recover a wide number of \emph{7}-manifolds of this type, analogously to what happens for Hermitian manifolds.

\item

Loosely speaking, $G_2$-manifolds are usually meant to be related to the Calabi-Yau manifolds, namely to Ricci-flat K\"ahler manifolds. Thus $G_{2}$-manifolds play a crucial role in compactifying \emph{11}-dimensional M-theory, analogously to the role of Calabi-Yau \emph{3}-folds in ten-dimensional string theory \cite{Gukov}. 

\item
The introduction of $G_2$ manifolds allows a breakdown of supersymmetry down to a smaller subgroup that only involves 1/8 of the original symmetry generators. This also entails compactifications of M-theory on $G_2$ manifolds to \emph{4}-dimensional theory with $\mathcal{N}=1$ supersymmetry, more suitable to make contact with eventual physical observations.

\end{itemize}

In what follows, we will consider the coassociative form $G$ --- in terms of which it is possible to reconstruct the metric --- as the field strength of a gauge potential, and write
\begin{eqnarray}
G=G_0+d\Gamma\,,
\end{eqnarray}
$\Gamma$ being a \emph{3}-form under which the membrane is charged. The $G_2$ manifold is equipped with the 3-form $\Phi$ and the dual 4-form $G=\star \Phi$ --- related to the metric, as mentioned above\footnote{Within the framework of a topological M-theory on G2, topological A,B-model branes and fields can be naturally unified. In particular, the 3-forms $\Phi$ and 4-forms $G$ combine fields of the A,B-models on the boundary with unit normal direction $dt$: $$\Phi={\rm Re}\Omega+k\wedge dt\,,$$
$$G={\rm Im} \Omega \wedge dt+\frac{1}{2}k\wedge k\,.$$
In other words, the A,B-models are interpreted as independent models only at perturbative levels: M-brane naturally couples the fields of the two models. The A-model is wrapped on Lagrangian cycles, in turn measured by 3-form, which is identified with the same holomorphic 3-form $\Omega$ in the B-model. $\delta \Omega$ in the B-model is the variation of a holomorphic 3-form on a Calabi-Yau 3-fold $X$. Similar considerations hold for $k$, which is the Kh\"aler form of the A-model while measuring the volume of holomorphic cycles in the B-model. In particular, A,B-model fields are canonically conjugate in the Hamiltonian reduction of topological M-theory on $X\times R$. The two models are connected by a S-duality. }.

\subsection{Action and Hamiltonian reformulation}
\noindent
We focus now on the topological M-theory on a 7D manifold $\mathcal{M}$ equipped with a real three form $\Phi$. The action is described by
\cite{Hitchin:2000jd,Hitchin:2001rw}
\be{HH}
I=\int_{\mathcal{M}}\sqrt{h(\Phi)}\,,
\ee
where 
$$\sqrt{h}h_{ab}=\Phi_{acd}\Phi_{bef}\Phi_{ghi} \epsilon^{cdefghi}\, .$$
In other words, the metric tensor can be completely rewritten in terms of a three form field $\Phi$. This is very much the same of what happens in LQG, where the densitized metric is cubic in the form field. Equivalently, the action can be rewritten in terms of the dual 4-form field $G$ instead of $\Phi$.

One can fix the cohomology 
class of $\Phi$: 
\be{PhiPhi}
\Phi=\Phi^{0}+d\beta\, , 
\ee
where $\beta$ is a two form. With such ``fixing'', which individuates a class of cohomologies, the action is invariant under gauge transformations, which are locally parameterized by a 1-form $\lambda$:
\be{beta}
\beta \rightarrow \beta'=\beta+d\lambda\,.
\ee

The Hitchin action can be recast as 
\be{IPHI}
I[g,\Phi]=\int_{\mathcal{M}}[\sqrt{g}-g^{ab}\sqrt{h}h_{ab}]\,,
\ee
providing the same equations of motion upon variation of the $g$ and $\Phi$ fields. We can focus on $\mathcal{M}=\Sigma \times R$, where $\Sigma$ is a 6D manifold. Fixing the time coordinate, we can define the canonical momenta
\be{PIi}
\pi=\frac{\delta I}{\delta \dot{\beta}_{ij}}\, ,
\ee
where the dot is the usual derivative on $R$ coordinates. The primary constraints are
generated by 
\be{pc}
\pi^{0i}=\frac{\delta I}{\delta \dot{\beta}}=0\, ,
\ee
while the Poisson algebra is 
\be{PA}
\{\beta_{ij}(x),\pi^{kl}(y)\}=\delta_{ij}^{kl}\delta^{6}(x,y)\, .
\ee
The Hamiltonian constraint can be written as --- see {\it e.g.} Ref.~\cite{Smolin:2005gu} --- 
\be{HC}
\mathcal{H}=\mathcal{K}-c \mathcal{V}\, ,
\ee
where $c$ is an adimensional constant, while
\be{KH}
\mathcal{V}=\tilde{\kappa}_{i}^{j}\tilde{\kappa}_{j}^{i}\, , 
\ee
\be{ktilde}
\tilde{\kappa}_{i}^{j}=\Phi_{ikl}\Phi_{mno}\epsilon^{klmnoj}\, ,
\ee
\be{Kaap}
\mathcal{K}=\pi^{ij} \pi^{kl} \pi^{mn} \epsilon_{ijklmn}\, .
\ee

Consequently, smearing against a test function $N$, the scalar constraint becomes 
\be{HN}
\mathcal{H}(N)=\int_{\Sigma} N\, (\pi^{ij} \pi^{kl} \pi^{mn} \epsilon_{ijklmn}-a\tilde{\kappa}_{i}^{j}\tilde{\kappa}_{j}^{i})\,, 
\ee
which  closes the Poisson algebra 
\be{HNHM}
\{\mathcal{H}(N),\mathcal{H}(M)\}=\int_{\Sigma}D_{j}\omega_{NM}^{j}\, ,
\ee
where 
\be{omegaa}
\omega_{NM}^{j}=18a(N\partial_{i}M-M\partial_{i}N)\pi^{ik}\tilde{\kappa}_{k}^{j}\, ,
\ee
$a$ being a numerical factor fixed to $1/4$ by the investigation of the Hamiltonian constraint.

\subsection{Gravity forms in \emph{3} and \emph{4} dimensions}
\noindent 
There exist in the literature examples of gravity forms theories in lower than \emph{7} dimensions. These theories are believed to belong to the quantum world, as can be still connected to topological M-theory, through dimensional reduction of topological M-theory. Below, we discuss few relevant cases for our discussion, namely theories of gravity forms in \emph{3} dimensions and \emph{4} dimensions. 

According to Ref. \cite{Dijkgraaf:2004te}, within the context of topological M-theory on a $G_{2}$ manifold, a metric theory can be reconstructed from the 3-forms $\Phi$ or dually from 4-forms $G=\star \Phi$. This amounts to say that the metric is not a fundamental field of this theory, but it can be reconstructed from the $\Phi$-field.

\subsection{BF theory, Chern-Simons and 3D gravity with cosmological constant}
\noindent 
The Einstein-Hilbert action with cosmological constant, which reads
\begin{eqnarray} 
\mathcal{S}_{\rm GR}=\int_{\mathcal{M}_3} \sqrt{-g} \, (E-2 \Lambda)
\,,
\end{eqnarray}
is a topological theory in \emph{3} dimensions. In order to introduce local degrees of freedom within the framework, one should indeed resort to higher order derivatives terms. This property of Einstein gravity in \emph{3} dimensions makes the problem of quantum gravity solvable, while dealing with the reformulations of the theory in terms of more explicit topological theories. As a matter of facts, the Einstein-Hilbert action rewrites in the first-order formalism as 
\begin{eqnarray} \label{3dl}
\mathcal{S}=\int_{\mathcal{M}_3} {\rm Tr}
\left( e\wedge F+\frac{\Lambda}{3} e\wedge e \wedge e \right),
\end{eqnarray}
with $F(A)=dA+A \wedge A$ the 2-form field-strength of an SU(2) gauge connection $A$
and $e$ the 1-form on $M$ valued in $SU(2)$. The metric is related to vielbiens $e_{a}^{i}$ as follows
\begin{eqnarray}
g_{ab}=-\frac{1}{2}{\rm Tr}(e_{a}e_{b})\,.
\end{eqnarray}
Action \eqref{3dl} can reformulated as a Chern-Simons gauge theory 
\begin{eqnarray}
S=\int_{M}{\rm Tr}\left(\mathcal{A}\wedge d\mathcal{A}+\frac{2}{3}\mathcal{A}\wedge \mathcal{A}
\wedge A\right)\, ,
\end{eqnarray}
in which $\mathcal{A}$ is the gauge connections of $SL(2,\mathbb{C})$ $(\Lambda<0)$ or $ISO(3)$ $(\Lambda=0)$ or $SU(2)\times SU(2)$ ($\Lambda>0$) in the Euclidean theory. 

\subsection{LQG in 3D }

Quantum 3D gravity can be quantized developing several different discrete models. In presence of non-vanishing cosmological constant $\Lambda$, a way is provided by the Turaev-Viro model: given a triangulation $\Delta$ of $M$, a quantum $6j$-symbol can be associated to each thetrahedron. One then obtains 
\begin{eqnarray} \label{TV}
TV(\Delta)=\left(-\frac{(q^{1/2}-q^{-1/2})^{2}}{2k}\right)^{V}\sum_{j_{e}}
\prod_{edges} [2j_{e}+1]_{q}\prod_{tetrahedra} (6j)_{q}\,,
\end{eqnarray}
$V$ being the total number of vertices in the triangulation, $[2j+1]_{q}$ denoting the quantum dimension of the spin $j$ representation of $SU(2)_{q}$ defined as 
\begin{eqnarray}
[n]_{q}=\frac{q^{n/2}-q^{-n/2}}{q^{1/2}-q^{-1/2}}\,.
\end{eqnarray}
The fundamental property of \eqref{TV} stands in its invariance from the triangulation 
scheme, which actually follows because of topological invariance:
\begin{eqnarray} 
TV(M)=TV(\Delta)\,.
\end{eqnarray}

\subsection{4D Gravity}

The action of the 4D self-dual sector of LQG reads 
\begin{eqnarray}
S=\int_{M^{4}}\Sigma^{k}\wedge F_{k}-\frac{\Lambda}{24}\Sigma^{k}\wedge \Sigma_{k}
+\Psi_{ij}\Sigma^{i}\wedge \Sigma^{j}\,.
\end{eqnarray}
Here $A^{k}$ is an $SU(2)$ gauge field with $F^{k}=dA^{k}+e^{ijk}A^{i}\wedge A^{j}$ and $\Sigma_{k}$ is a $SU(2)$ triplet of 2-forms fields --- $i,j,k=1,2,3$; $\Psi_{ij}$ is a scalar field on $M$ which is a symmetric representation of $SU(2)$. Varying the action with respect to $\Psi_{ij}$, we can derive 
\begin{eqnarray}
\Sigma^{(i}\wedge \Sigma^{j)}-\frac{1}{3}\delta^{ij}\Sigma_{k}\wedge \Sigma^{k}=0\,.
\end{eqnarray}
Such a condition implies that $\Sigma^{k}$ can
be re-expressed in terms of the vierbein 
\begin{eqnarray}
\Sigma^{k}=-\eta_{ab}^{k}e^{a}\wedge e^{b}\, ,
\end{eqnarray}
where $e^{a}$ is the vierbien 1-forms on $M^{4}$, 
$a=1,...,4$ and $\eta_{ab}^{k}$ is the 't Hooft symbol
$$\eta_{ab}^{k}=\epsilon_{ab0}^{k}+\frac{1}{2}\epsilon^{ijk}\epsilon_{ijab}\,.$$
More explicitly 
\begin{eqnarray}
\Sigma^{1}=e^{1}\wedge e^{2}-e^{3}\wedge e^{4}\,,
\end{eqnarray}
\begin{eqnarray}
\Sigma^{2}=e^{1}\wedge e^{3}-e^{4}\wedge e^{2}\,,
\end{eqnarray}
\begin{eqnarray}
\Sigma^{3}=e^{1}\wedge e^{4}-e^{2}\wedge e^{3}\,.
\end{eqnarray}
As well known, the vierbein is in turn related to the metric by the relation  
\begin{eqnarray}
g=\sum_{a=1}^{4}e^{a}\otimes e^{a}\,.
\end{eqnarray}

The two-forms $\Sigma^{k}$ are self-dual with respect to the metric $g$, {\it i.e.} $\Sigma^{k}=*\Sigma^{k}$. One can also rewrite the metric directly in terms of $\Sigma$, finding 
\begin{eqnarray}
\sqrt{g}g_{ab}=-\frac{1}{12}\Sigma_{aa_{1}}^{i}\Sigma_{ba_{2}}^{j}\Sigma_{a_{3}a_{4}}^{k}\epsilon^{ijk}\epsilon^{a_{1}a_{2}a_{3}a_{4}}\,.
\end{eqnarray}

\subsection{Reducing Topological M-theory to Gravity}
\noindent
We address now local models of a complete 7-manifold $X$ obtained as a m-dimensional vector bundle\footnote{The relation between 7D Hitchin-like theories and the chiral formulations of 4D gravity was addressed by Krasnov in Ref.~\cite{Krasnov:2016wvc}. The dimensional reduction on $S^3$ of the topological theory with Lagrangian $CdC$ corresponds to a 4D $BF$ theory on $SU(2)$ principle bundle, with cosmological constant term related to the radius of $S^3$.} on an n-dimensional cycle $M$, such that $m+n=7$. Local gravitational modes induce a lower-dimensional gravity on $M$. The equations of motion of topological M-theory 
\begin{eqnarray}
d\Phi=0\,,
\end{eqnarray}
\begin{eqnarray}
d_{\star\Phi}\Phi=0\, ,
\end{eqnarray}
lead to the equations of motion of the p-form fields on $M$, in turn interpreted as topological gravity equation of motion  on $M$. 

The cases in which $n=3$, $m=4$ and $n=4$, $m=3$ can be discussed in the same construction framework. $\Phi$ can be decomposed as a combination of vielbein components $e^i$. 

The equation $d\Phi=0$ is equivalent to the equations of motion of 3D gravity for 3D fiber, namely 
\begin{eqnarray}
&& de=-A\wedge e-e\wedge A\,, \nonumber \\
&& dA=-A\wedge A-\Lambda e\wedge e\,,
\end{eqnarray}
which are precisely equivalent to the equations of motion of 3D Chern-Simons gravity, {\it i.e.}
\be{dA}
d\mathcal{A}+\mathcal{A}\wedge \mathcal{A}=0\, .
\ee
The latter also fulfills $d_{\star\Phi}\Phi=0$. The $\Phi$-field can be rewritten as a combination of vielbein as
\be{PhiA}
\Phi=Ae^{123}+Be_{i}\wedge \Sigma^{i}\,,
\ee
where 
\be{Sigmau}
\Sigma^{1}=\alpha^{12}-\alpha^{34}, \cdots
\ee
and $\alpha_{i}$ are 1-forms in the fiber direction 
\be{alpha}
\alpha^{i}=D_{A}y^{i}=dy^{i}+(Ay)^{i}\,.
\ee

Now let us consider the reduction to the 4D gravity.  In this case, we can decompose $\Phi$ as
\be{Phialphap}
\Phi=\alpha^{123}+\alpha^{1}\wedge \Sigma^{1}
+\alpha^{2}\wedge \Sigma^{2}+\alpha^{3}\wedge \Sigma^{3}\,.
\ee

\subsection{Quantization of the topological M-theory.}
\noindent
In Ref.~\cite{Smolin:2005gu} Smolin suggested a quantization scheme, defining the holonomy of the 1-form $\beta$ and its conjugate variables, respectively, as
\be{TS}
H[S]=e^{\int_{S}\beta}
\ee
and the momentum flux operators 
\be{PiA}
\Pi[A]=\int_{A}\pi^{*}\,.
\ee
The Poisson brackets can be then recovered as
\be{TSA}
\{H[S],\Pi[A]\}=I[S,A]H[S]\, ,
\ee
where $I[S,A]$ is the intersection number of the surfaces $S,A$.

This allows to define networks $\Gamma$ on the two surfaces, with their relative Hilbert states such that 
\be{Psi}
\langle \Gamma|\Psi\rangle=\Psi(\Gamma)\, .
\ee
 
\subsection{Dimensional reduction}

\noindent
We consider now the semiclassical limit of the topological M-theory. The 11 dimensional supergravity action 
\be{MMM}
\int_{\mathcal{M}}da\wedge da \wedge a
\ee
can be obtained. This action corresponds to a higher-dimensional Chern-Simons theory 
\cite{Ling:2000dk}. We can define a canonical momentum for $a$, for which $\Pi^{*}=a \wedge da$. Equation (\ref{MMM}) is derived consistently taking the connection, the frame field and gravitinos to zero. This allows to consider only the action of the 3-form. 

Moving then from the 11-dimensional action in (\ref{MMM}), we can consider the dimensional reduction
$$\mathcal{M}^{11}=R\times \Sigma_{6} \times S^1 \times R^{3}\,.$$ 
This amounts to a dimensional reduction of the momenta specified by
\be{pip}
\pi^{*}=\int_{R^{3}}\Pi^{*}\, ,
\ee
\be{betpi}
\{\beta_{ij}(x),\pi_{klmn}^{*}(y)\}=\int_{S^{1}} d\theta \int d^{3}x^{\alpha\beta\gamma}\{a_{\theta ij},\Pi^{*}_{klmn\alpha\beta\gamma} \}=\epsilon_{ijklmn}\delta^{6}(x,y)\,.
\ee
The canonical degrees of freedom of topological M theory are obtained from the dimensional reduction of 11-D supergravity. 

\section{Aharonov-Bohm effect in LQG}

In this section, we will discuss a reformulation, suggested in Ref.\cite{Bianchi:2009tj}, of states and scalar products of LQG in terms of non-trivial holonomies enclosing defects. One can start from a 3-manifold $\Sigma$ with a network of defect-lines. To a locally-flat connection on the 3-manifold one can associate a non-trivial holonomy, as in the electromagnetic Aharanov-Bohm effect. Quantizing the theory, Bianchi obtained a scalar product that is the same used in LQG. 

We consider a flat connection in $\Sigma'=\Sigma-l$, where $l$ is a defect line, and then the holonomy encircling this line. The induced metric on $\Sigma$ is $q_{ab}(x)$, which allows to choose the Coulomb-gauge as $\chi=q^{ab}\partial_{a} A_{b}$. The line is fixed along the z-axes in the Euclidean metric. Considering the gauge fixing condition
\be{chi}
\chi^{i}=\partial^{a}A_{a}^{i}=0\,,
\ee
$A_{a}^{i}$ reads
\be{Aai}
A_{a}^{i}=\frac{f^{i}}{2\pi}\alpha_{a}(x)\, ,
\ee
where $f_i$ stands for the flux of the magnetic field through the defect line, and
\be{alpha}
\alpha_{a}(x)=\left(-\frac{y}{x^{2}+y^{2}},\frac{x}{x^{2}+y^{2}},0\right)\, .
\ee
The associated holonomy along the loop $\gamma$ is 
\be{hgammaA}
h_{\gamma}[A]={\rm exp}\left[i\left(\int_{\gamma}\alpha_{a}dx^{a}\right)\frac{f^{i}}{2\pi}\tau_{i}\right]\, ,
\ee
while the related non-abelian magnetic field reads 
\be{Biii}
B_{i}^{a}\equiv \frac{1}{2}\epsilon^{abc}F_{bc}^{i}=\int_{l}ds f_{i} \dot{x}^{a}(s)\delta^{(3)}(x-x(s))\, .
\ee
The flux of the magnetic field through the surface $S$ punctured by the curve $\gamma$ reads 
\be{MFF}
\mathcal{F}_{i}[B,S]=\int_{S} B_{i}^{a}(X(\sigma))\epsilon_{abc}\frac{\partial X^{b}}{\partial \sigma^{1}}\frac{\partial X^{c}}{\partial \sigma^{2}}d\sigma^{1}d\sigma^{2}
\ee
$$=\int_{l} ds \int_{S}d\sigma^{1}d\sigma^{2}f_{i}\epsilon_{abc}\dot{x}^{a}(s)\frac{\partial X^{b}}{\partial \sigma^{1}}\frac{\partial X^{c}}{\partial \sigma^{2}}\delta^{(3)}(X(\sigma)-x(s))\,, $$
which is just equal to $f_i$. This is analogous to the problem of a cylindrical solenoid in electromagnetism.

The related moduli space is 
\be{Phiiop}
\{ f^{i}\in S_{3}\}/SU(2)=\{\phi \in [0,2\pi]\}\, , 
\ee
where $\Psi^i$ depends only on the moduli $\phi$, while is invariant under global $SU(2)$ rotations.

In this framework, the scalar product of states depending by the moduli space can be put in correspondence with the LQG scalar product of holonomy states in the Hilbert space $\mathcal{K}'$. We then find
\be{lfigr}
\langle g|g' \rangle=\int_{\mathcal{A}_{f}/\mathcal{G}}\mathcal{D}[A]\bar{\Psi}_{g}[A]\Psi_{g'}[A] =\int_{\mathcal{N}}\prod_{r}dm_{r}\, J(m_r) \Delta_{FP}(m_r) \bar{g}(m_r) \,g'(m_r)\,,
\ee
where $\{m_r\}$ denotes the moduli space, $J$ stands for the Jacobian and $\Delta_{FP}$ the Faddeev-Popov determinant. In order to prove this equivalence, we perform explicitly the computation for the case of one line defect. 
The Jacobian in spherical coordinates reads  
\be{Jphiphi}
J(\phi)=\phi^2 \, , 
\ee
which is associated to
\be{d2p}
d^{2}\Phi=\phi^{2}d\phi d^{2}v^{i}\,.
\ee

The Faddeev-Popov term $\Delta_{FP}$ is given by the determinant of the operator
$K$, which is the derivative of the gauge fixing condition $\chi^i=0$ with respect to the gauge parameter, namely
\be{Kdeltsa}
K=-\delta_{ij}\Delta-\epsilon_{ijk}\frac{\Phi^{k}}{2\pi}\alpha^{a}\partial_{a}\,.
\ee
Its eigenvalues are
\be{lambdaa}
\lambda_{n}=n^{2}+n\frac{\phi}{2\pi}\, ,
\ee
where $n=\pm 1,\pm 2,\dots$ (twice degenerate). 
The (regularized) Faddeev-Popov determinant can be then cast as
\be{DeltaFP}
\Delta_{FP}(\phi)=c\frac{{\rm Det} K(\Phi^{i})}{{\rm Det}K(0)}=c\frac{\prod_{n=1}^{\infty}(\lambda_{n}(\phi))^{2}(\lambda_{-n}(\phi))^{2}}{(\lambda_{n}(0))^{2}(\lambda_{-n}(0))^{2}}
\ee
$$=c\prod_{n=1}^{\infty}\left(1-\left(\frac{\phi}{2\pi}\right)^{2}\right)^{2}=c\left( \frac{\sin \phi/2}{\phi/2}\right)^{2}\, ,$$
where the constant $c$ is fixed by imposing normalization to $1$. Using these expressions we obtain 
\be{lfigo}
\langle g|g'\rangle=\frac{1}{\pi}\int_{0}^{2\pi} \sin^{2}(\phi/2) \,\bar{g} \, g'\,.
\ee
This can be compared to the scalar product of LQG. A natural choice in LQG is the Haar measure on the links of the graphs, which reads 
\be{ligraphs}
\langle \eta|\zeta\rangle=\int_{\mathcal{A}_{l}/\mathcal{G}}\mathcal{D}[A]\bar{\Psi}_{\Gamma,\eta}[A]\bar{\Psi}_{\Gamma',\zeta}[A]
\ee
$$=\int_{SU(2)^{L}}\prod_{l=1}^{L}d\mu_{H}(h_{l})\bar{\eta}(h_{1},...,h_{L})\zeta(h_{1},...,h_{L})\, ,$$
in terms of the class of graphs $\Gamma'$ dual to the cellular decomposition. Using the Peter-Weyl theorem with such choice of the scalar product, the spin-network states, with graph $\Gamma'$, provide an orthonormal basis of the Hilbert space $\mathcal{K}'$. The spin-network basis provide cylindrical functions $\eta(h_{1},...,h_{L})$, the holonomies of which are labeled by $SU(2)$ representations. To every node of the spin-network states are assigned intertwiners that realize an invariant map onto the tensor product of the representations. In particular, we can recognize that 
\be{etajn}
\eta_{j_{1}j_{n}}(h_{1},...,h_{L})=\left(\bigotimes_{n}v_{in}\right)\left(\bigotimes_{\gamma_{l}}D^{(j_{l})}(h_{L})\right)\,,
\ee
where $n \in \Gamma'$, $\gamma_{l}\in \Gamma'$, in such a way to have orthonormality of the spin-network states
\be{etalkl}
\langle \eta_{j_{1}i_{n}}|\eta_{j'_{l}i'_{n}}\rangle=(\prod_{l} \delta_{j_{l}j_{l}'})(\prod_{n}\delta_{i_{n}i_{{n}'}})\,.
\ee

The states 
\be{finalstate}
\Psi_{\gamma,\eta}[A]=\eta(h_{\gamma}[A])
\ee
are associated by a complex-values function $\eta$ on $SU(2)$, and by the homotopy class $[\gamma]$ of loops closing one time the defect $l$. The scalar product can be cast in terms of the Haar measure on $SU(2)$. In particular, if we define $f(\phi)=\eta(e^{i\phi\tau_{3}})$, we obtain $\langle \eta|\zeta \rangle=\langle g|g'\rangle$.

\begin{figure}[t!!]
\begin{center}
\vspace{1cm}
\includegraphics[width=6cm,height=7cm,angle=0]{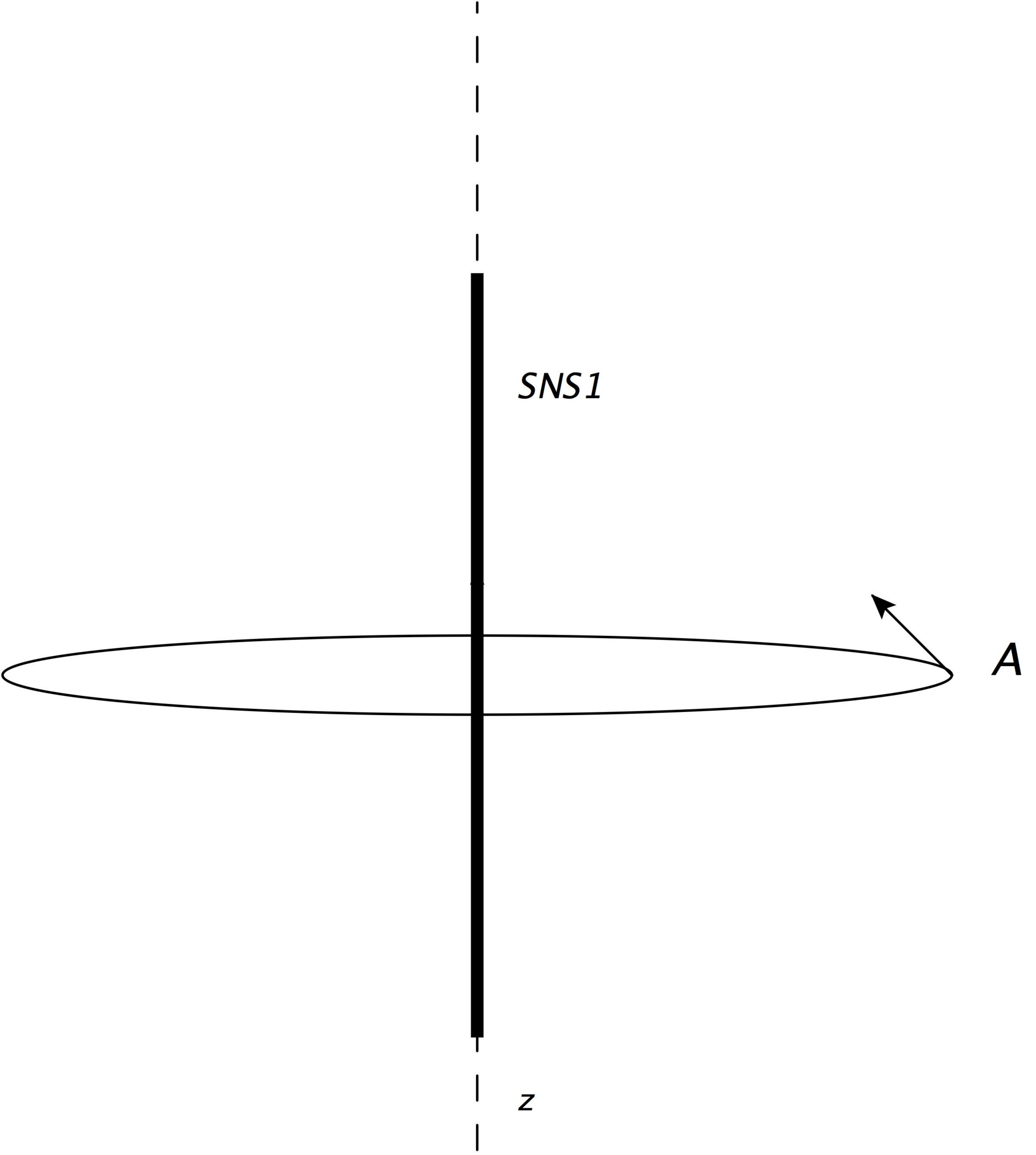}
\caption{ The presence of the space-like fundamental string is associated to a non-trivial holonomy of self-dual variables.
 } \label{fig:1}
\end{center}
\end{figure}

\subsection{Generalization of the argument}

One can generalize the example of one line defect into the case of a network $\mathcal{S}$ of curves in $\Sigma$ \cite{Bianchi:2009tj}. Let us define a locally-flat connection $A(x)$. The holonomies of these connections are not trivial, since they encircle the net of defects. Let us consider the space of locally flat connections, modulo the gauge transformations, which we call $\mathcal{A}_{f}/\mathcal{G}$. Now, the configuration in such a space  corresponds to a homomorphism from $\pi_{1}(\Sigma-\mathcal{S})$ into $G$, cosetting gauge transformations. This defines the moduli space $\{m_{r}\}$ contained in 
 \be{N}
 \{m_{r}\}\equiv\mathcal{N}\equiv {\rm Hom}\{\pi_{1}(\Sigma-\mathcal{S},G)/G\}\, .
 \ee

In this generalized set-up, the states can be put in correspondence with functions of the moduli,
\be{PsiGammae}
\Psi_{\Gamma,\eta}[A^{m_{r},g}]=f(m_{1},...,m_{r})\, ,
\ee
where 
\be{Psi14}
\Psi_{\Gamma,\eta}[A]=\eta(h_{\gamma_{1}}[A],...,h_{\gamma_{L}}(A))\, 
\ee
and $\eta$ is a complex function valued in $SU(2)$, {\it i.e.} 
\be{etatb}\eta:SU(2)^{L}\rightarrow C\, .
\ee
The scalar products of LQG and the moduli functions are in correspondence by means of
\be{lahl}\langle f|g\rangle =\int_{\mathcal{A}_{f}/\mathcal{G}}\mathcal{D}[A]\bar{\Psi}_{f}[A]\Psi_{g}[A]=\int_{\mathcal{N}}d\mu(m_{r})\bar{f}(m_{1},...,m_{R})g(m_{1},...,m_{R})\,.
\ee

\section{SM-branes and S-branes}
\noindent
Space-like branes, or S-branes, are very similar to ordinary branes, but completely localized in space-like coordinates, {\it i.e.} they have not time-like coordinates, which implies that they are unstable. Sp-branes must be contained in string theory, appearing in correspondence of a tachyonic kink field localized along the time-direction with a tachyonic potential interpolating two minima of two unstable (p+1)-brane --- among the many references on this subject, see {\it e.g.} Refs.~\cite{Sen:2002nu,Sen:2002in,Gutperle:2002ai,Strominger:2002pc,Hashimoto:2002sk,Ohta:2003uw,Strominger:2001pn,13}. 

The existence of S-branes is believed to play an important rule within the context of dS/CFT correspondence \cite{Strominger:2001pn}. Strominger has also conjectured that in the Large N limit of the number of S-branes, they may be holographically dual to interesting closed string gas cosmologies \cite{Strominger:2002pc}. 

As for the ordinary instantonic branes, there is a huge zoology of possible S-branes. Very similarly to M-branes --- they are M2-branes and M5-branes --- M-theory predicts SM2 and SM5-branes. From SM2 and SM5-branes, we can construct several different branes in string theories from direct compactification, double compactifications and duality transformations. 

From SM2-branes and SM5-branes SD2, SNS5, SNS1 and SD4 are obtained by direct or double dimensional reduction \cite{13}. On the other hand, a SDp-brane in IIA superstring is related to a NS-brane and fundamental strings in IIA superstring, by using S-duality.

Moreover, systems of multicharged S-branes and bound states can be constructed as in the case of D-branes --- for example like D1/D5 in AdS/CFT. In particular, possible bound state systems are: {\it i)} $SDp/SD(p-2)$-brane solution with $p\geq 2$ (like SD2/SD0-brane); {\it ii)} $SDp/SD(p-4)$-brane solutions with $p\geq 4$ (like $SD4/SD0$); {\it iii)} $SDp/SD(p-6)$-branes, the only natural one (like $SD6/SD0$-brane).

\subsection{S-branes instabilities and particle productions}
\noindent 
Sen argued that in the worldsheet boundary, $\sinh X^{0}$ --- $X^{0}$ being the time-like coordinate --- gives an exact conformally invariant boundary sinh-Gordon field theory
\cite{Sen:2002nu,Sen:2002in}. This happens in the limit of $g_{s}\rightarrow 0$, where quantum effects of closed strings are suppressed. 

Away from $g_{s}=0$, one could expect that S-branes decay into closed strings. Strominger and Gutperle have studied the case of $e^{X_{0}}$, which corresponds to a boundary Liouville theory with negative norm boson \cite{Gutperle:2002ai,Strominger:2002pc}. This may be formally obtained from the Sen model by taking the location $a$ of the brane in the past infinity, and rescaling the interaction strength. In this limit, the energy of the s-brane is converted into a pressureless tachyonic dust, confined among D-branes.

The equation of motion is a Klein-Gordon equation with a time-dependent mass. For slowly varying $m$, the solution is like 
\be{eet}
e^{\pm i E(t)t},\,\,\,\, \qquad E^{2}=m^{2}+p^{2}\,.
\ee
Now, in the solutions of the Klein-Gordon equation, the incoming modes
\be{incm}
t\rightarrow -\infty ,\,\,\,\, \qquad \phi_{p}^{IN}\sim e^{-i\omega t+p\cdot x}\, 
\ee
have both negative and positive frequency parts in the far future, {\it i.e.} particles are produced. In particular, as a limit of the Klein-Gordon solutions to the equations of motion --- proportional to a combination of Henkel functions --- one finds 
\be{phips}
\phi_{p}^{OUT}\rightarrow e^{-t/2-ie^{t}+ip\cdot x}\, ,
\ee
while the energy is going as 
\be{Ero}
E(t)=|\dot{\phi}^{OUT}|^{2}\sim e^{t}\sim m(t)\, .
\ee
This leads to the Hogedorn divergence. The energy from the decaying brane reads 
$$\int N_{\omega}dE_{\omega}\, ,$$
being calculated over all the open string modes with energy $\omega$ and density of states $N_{\omega}$. The differential energy $dE_{\omega}$ stands for the expectation value of the outgoing energy in open string modes. Since $E(t)\sim m(t)$, one tries a divergence --- the Hagedorn divergence --- in the integral. In fact, the differential energy reads 
\be{deom}
dE_{\omega}=\frac{d^{p}p}{(2\pi)^{p}}\frac{e^{2\pi X^{0}T_{Ha}}}{(e^{\omega/T_{Ha}-1})}\, ,
\ee
where $T_{Ha}$ is the Hagedorn temperature,
\be{THa}
T_{Ha}=\frac{1}{4\pi \sqrt{\alpha'}}\,.
\ee
At large values of $\omega$, the number density can be approximated as 
\be{Nom}
N_{\omega} \sim \omega^{-a}e^{\omega/T_{H}}\, , 
\ee
$a$ being the number of non-compact directions transverse to the brane, and 
\be{sEml}
dE_{\omega}\sim \omega^{p-1}e^{-\omega/T_{H}}e^{2\pi X^{0}T_{H}}d\omega\,.
\ee
In other words, open strings receive an infinite energy from high-energy modes
--- the brane produces open-strings at the Hagedorn temperature. 

The same conclusions can be reached from the worldsheet side, in the {\it minisuperspace approximation}, {\it i.e.} the quantization of the open strings in the minisuperspace approximation for the zero mode $X^{0}$. This approximation is very much similar to the ordinary bulk Liouville theory \cite{L1,L2}. The (bosonic part of the) wolrdsheet action of an open strings on an unstable D-brane reads
\be{XXs}
-\frac{1}{4\pi \alpha'} \int d\tau d\sigma \sqrt{-\gamma}\gamma^{ab}\partial_{a}X^{\mu}\partial_{b}X_{\mu}-\frac{1}{8\pi}\int d\tau \sqrt{-h}T(X)\, ,
\ee
where $T$ is the background tachyonic field --- for superstrings $T\rightarrow T^2$, and the stability can be ensured.

In a minisuperspace approximation, the zero mode is treated as independent from higher oscillatory modes. In Refs.~\cite{Gutperle:2002ai,Strominger:2002pc}, the case considered is an exponential tachyonic profile, namely 
\be{THJK}
T(X)\equiv T(X^{0})=e^{X_{0}/\sqrt{\alpha'}}\,.
\ee

For $X_{0}\rightarrow -\infty$, the tachyon is at the top of its potential. The closed string vacuum is reached in the far future, at which open strings become infinitely massive and a continuos spectrum is reached. Open string masses are in the exponential form 
\be{mDL}
m^{2}(X^{0})=m_{0}^{2}+\frac{1}{4\pi \alpha'}e^{X^{0}/\sqrt{\alpha'}}\,.
\ee

\subsection{S-branes and generation of fundamental strings}
\noindent 
The SD-brane action is a Dirac-Born-Infeld action for Euclidean world-volumes. As mentioned above, this action is related to the presence of a time-like tachyonic condensate. The time-like tachyonic action reads 
\be{SS}
S=-\int d^{p+2}x V(T)\sqrt{1+(\partial_{\mu} T)^{2}}\, . 
\ee
This action is provided with BIonic-type solutions that are associated to the appearance of a fundamental string \cite{Hashimoto:2002sk}. In particular, the SD-brane action reads  
\be{detAA}
\sqrt{{\rm det}(\delta_{\hat{\mu}\hat{\nu}}-\partial_{\hat{\mu}}X^{0}\partial_{\hat{\nu}}X^{0}+\partial_{\hat{\mu}} A_{p+1}\partial_{\hat{\nu}} A_{p+1})}\, , 
\ee
which has a Sp-brane spike solution reading  
\be{Xz}
X^{0}=A_{p+1}=\frac{C_{p}}{r^{p-2}}\, ,
\ee
where $r^{2}=\sqrt{x_{1}^{2}+...+x_{p}^{2}}$ is the radial coordinate along the Euclidean worldvolume. 

The S-brane action can be rewritten as 
\be{SSz}
S=S_{0}\int dx^{0}d^{p}x\sqrt{-1+E_{p+1}^{2}+\dot{r}^{2}}\, ,
\ee
which is associated to a Hamiltonian density 
\be{HH}
\mathcal{H}=\frac{S_{0}}{\sqrt{-1+E^{2}+\dot{r}^{2}}}\,. 
\ee
Imposing the quantization condition on the conjugate momenta
\be{P}
\int d^{p-1}x P_{E}=n\, ,
\ee
where 
\be{PEE}
P_{E}=\frac{S_{0}E}{\sqrt{-1+E^{2}+\dot{r}^{2}}}\, , 
\ee
the Hamiltonian $\mathcal{H}=P_{E}/E$ becomes 
\be{HHH}
\int dx^{p}\mathcal{H}=\frac{n}{2\pi \alpha'}\int dx^{p+1}\, ,
\ee
which corresponds to the Hamiltonian of $n$ static fundamental strings with fundamental string tension ${\bf T\sim \alpha'^{-1}}$.

\subsection{S-branes long-range interactions }
\noindent 
At the perturbative level, Sp-branes generate a long-distance potential of closed strings, which amounts to the emission of gravitons and dilatons. The boundary state --- considering the bosonic strings sector --- of the Sp-brane has a spatial component that reads
\be{BB}
|B\rangle_{\vect{X}}=\frac{T_{p+1}}{2}\delta^{8-p}(\vect{x})\, {\rm exp}\left(-\sum_{n=1}^{\infty}S_{ij}a_{-n}^{i}\tilde{a}_{-n}^{i}\right)\, , 
\ee
where $T_{p+1}$ is the $SD(p+1)$-brane sector and 
\be{Sijk}
S_{ij}=(\delta_{AB},-\delta_{ab})\, , 
\ee
with $A,B$ and $a,b$ corresponding to Neumann and Dirichlet directions respectively. 

The total source state for gravitons/dilatons undergoes the expansion 
\be{BBB}
|B\rangle=\frac{T_{p+1}}{2}\delta^{8-p}(\vect{x}_{\perp})\left[-f_{1}(X^{0})S_{ij}a_{-1}^{i}\tilde{a}_{-1}^{j}+f_{2}(X^{0})S_{ij}a_{-1}^{0}\tilde{a}_{-1}^{0} \right]|0\rangle+ \dots \,,
\ee
with
\be{fu}
f_{1}(X^{0})=\frac{1}{1+e^{X^{0}}\sin \pi \lambda}+\frac{1}{1+e^{-X^{0}}\sin \pi \lambda}-1\, , 
\ee
and 
\be{fd}
f_{2}(X^{0})=1+\cos 2\pi \lambda-f(X^{0})\, , 
\ee
having taken the time component expansion as
\be{BBBB}
|B\rangle_{X^{0}}=f_{1}(X^{0})|0\rangle+a_{-1}^{0}\tilde{a}_{-1}^{0}f_{2}(X^{0})|0\rangle+ \dots 
\ee
and resorted to the expansion in (\ref{BB}). \\

The massless part of the closed string reads 
\be{CC}
|C\rangle=\frac{T_{+1}V_{p+1}}{2}\int dt\Delta(X;0,t)[f_{2}(t)a_{-1}^{0}\tilde{a}_{-1}^{0}-S_{ij}f_{1}(t)a_{-1}^{i}\tilde{a}_{-1}^{j}]|0\rangle+... \,,
\ee
where $V_{p+1}$ is the spatial volume of the s-brane. The closed strings creation operator product expansion for
\be{Jmunu}
J^{\mu\nu}(k)=\langle 0;k|a_{1}^{\mu}\tilde{a}_{1}^{\nu}|C\rangle\,  
\ee
reads 
\be{JJ}
J_{00}(x)=-\frac{T_{p+1}V_{p+1}}{2}\frac{1}{4\pi r}\left[\frac{1}{1+e^{t-r}\sin \pi \lambda}+\frac{1}{1+e^{-t-r}\sin \pi \lambda}-2-\cos \pi \lambda \right]\, ,
\ee
\be{JJJ}
J_{ij}(x)=-\frac{T_{p+1}V_{p+1}}{2}\frac{S_{ij}}{4\pi r}\left[\frac{1}{1+e^{t-r}\sin \pi \lambda}+\frac{1}{1+e^{-t-r}\sin \pi \lambda}-1\right]\,.
\ee
In the large radius $r$ expansion, equations (\ref{JJ}) and (\ref{JJJ}) behave as 
\be{JJa}
J_{ij}\rightarrow -C_{p}\frac{S_{ij}}{r^{8-p}}\, , 
\ee
\be{JJJb}
J_{00}\rightarrow C_{p}\frac{\cos 2\pi \lambda}{r^{8-p}}\, ,
\ee
\be{Cp}
C_{p}=\frac{T_{p+1}V_{p+1}}{4}\pi^{\frac{p-8}{2}}\Gamma\left(\frac{8-p}{2} \right)\, . 
\ee

The dilatons' and gravitons' emissions correspond to ``annulus diagrams'', {\it i.e.} to the emission of closed strings from the (S)Dp-brane. Through the annulus diagrams, (S)Dp-branes can interact through the exchange of a closed string. This process can be calculated as a tree-level diagram in perturbation theory, in the low energy limit. However, the interaction in non-perturbative regime is generically impossible to be calculated, since it requests the knowledge of all the orders of perturbation theory.

\section{Correspondence between SM-brane foam and spinfoam}
\noindent 
The line defect introduced in Sec.~3 can be interpreted as a soliton charged in the self-dual gravitational gauge group --- see Fig.1 . Since the scalar constraint was not yet solved in full generality, we consider space-like solitons not propagating in the time direction. They can be either space-like strings or holonomies around a circular solenoid that take a circular path orthogonal to it. 

Our conjecture is that the line defects correspond to compactified SM-branes into space-like fundamental strings, {\it i.e.} space-like NS1-branes \footnote{In standard Dp-brane notation, it should be more correct to call it SNS0-branes.
Here, we use the convention Space-like NSp-brane in order to describe branes with $p$ 
instead of $p+1$ space-like directions. 
}. Fundamental S-strings are serious candidates to be considered in order to instantiate this correspondence. These are charged indeed with respect to the $\Phi$-field and consequently with respect to self-dual gravitational potential $A_{\mu}^{i}$, which contains the self-dual gravitational algebra structure --- see {e.g.} Sect.~2.6 . In particular, taking an holonomy $h_{\gamma}[A_{\mu}^{i}]$ encircling 
the fundamental S-strings, the magnetic flux on a surface $S$ punctured by the curve $\gamma$ is non-zero because essentially it surrounds a localized magnetic field from $G=\star \Phi$. 

There is a possible issue for this correspondence's framework: different S-branes networks can correspond to the same spin-networks and {\it viceversa}. In other words, the correspondence can be established from classes of S-branes networks to classes of LQG states. Nonetheless, the classes' correspondence is enough to guarantee that every possible S-branes systems have a proper state in the LQG Hilbert space.

\subsection{S-branes, coarse graining and flowers}

\begin{figure}[t!!]
\begin{center}
\vspace{1cm}
\includegraphics[width=12cm,height=4cm,angle=0]{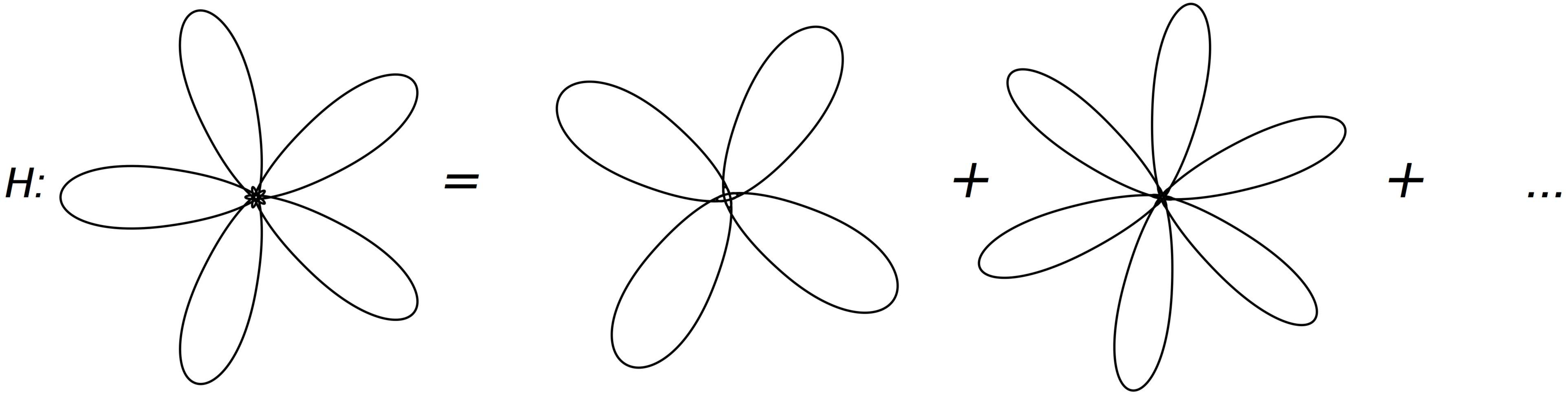}
\caption{ The scalar constraint applied to a flower diagram: the result is a quantum super- position of flowers with a removed and an added petal. This amounts to creating or annihilating a space-like fundamental strings. } \label{fig:1}
\end{center}
\end{figure}

\noindent
An interesting class of S-strings systems corresponds to {\it holonomy flowers}. Flowers are obtained from shrinking homotopy trees --- every possible paths without loops in the spin-network \cite{Livine:2013tsa,Livine:2013gna,Charles:2016xwc}. Each flower represents a class of S-strings --- S-strings encircled inside the petals. 

In particular, flowers correspond to a single vertex together with N-number of loops attached to it. The spin-network states supported on these graphs is associated to an intertwiner $\mathcal{I}$ with the tensor product 
\be{bigo}
\bigotimes_{l=1}^{N}(\mathcal{V}^{k_{l}}\otimes \bar{\mathcal{V}}^{k_{l}})\,,
\ee
where each loop is labelled by a spin $k_l$, with ${l=1,..,N}$. 

Associated states are gauge-invariant combinations of N group elements, {\it i.e.}
\be{Statet}
\Psi(h_{1},...,h_{N})=\Psi(gh_{1}g^{-1},...,gh_{N}g^{-1})\, .
\ee
The resulting Hilbert space, equipped with the Haar measure on $SU(2)$, is 
\be{HN}
\mathcal{H}_{N}=L^{2}\left(SU(2)^{N}/AdSU(2) \right)\, ,
\ee
with the basis 
\be{PsiO}
\Psi^{\{k_{l},\mathcal{I}\}}(\{k\}_{l=1,..,N})=\langle h_{l}|k_{l},\mathcal{I}\rangle={\rm Tr}\left[\mathcal{I}\otimes \bigotimes_{l=1}^{N}D^{k_{l}}(h_{l}) \right] \,,
\ee
having taken the trace over the tensor product (\ref{bigo}). \\

Flowers diagrams turned out to be particularly useful in the LQG coarse-graining procedure. Within the context of the S-branes reinterpretation of the spinfoam, coarse-graining is motivated in the limit in which UV degrees of freedom of M-theory are not fully excited and they can be integrated out --- this is very reminiscent of the renormalization group approach in condensed matter and quantum field theory. 

The coarse-graining approach seems also to suggest how to treat higher dimensional branes with respect to fundamental 
space-like strings. Higher dimensional branes are heavier and  can be thought as integrated out. The same procedure can be proposed, in many other cases, for higher dimensional branes in correspondence of nodes. Taking into account also these heavier degrees of freedom would complicate very much the dynamics of the system. Thus we suggest to use such approximation, dubbing it {\it light S-branes coarse graining}.

\subsection{Interpretation of the Scalar constrains}
\noindent
From the perspective of LQG, the action of the scalar constraint operator on flowers' states corresponds to writing a superposition of states in which a petal is either removed or added. The action of the Hamiltonian operator on the state can be cast as 
\be{HS}
H|S\rangle =\sum_{n\in S}N_{n}\sum_{l,l',l'',r}\sum_{\epsilon',\epsilon''=\pm}H_{n,l',l'',\epsilon',\epsilon''}D_{n,l',l'',r,\epsilon',\epsilon''}|S\rangle\, .
\ee
As renown, $H_{n,l',l'',\epsilon',\epsilon''}$ acts on the space of n-valent intertwiners at the node, and generate or destroy a loop of the flower. This has a nice physical reinterpretation from the M-theory perspective. The disappearance of a petal corresponds to the disappearance of a SNS1-string, {\it i.e.} it can be reinterpreted as a decay process into open or closed strings. This kind of processes is very much expected also from considerations summarized in Sec.~4, from the perturbative string theory approach. \\

On the other hand, the process of creation of a fundamental S-string corresponds to the nucleation of S-strings from other branes. The nucleation rate can be estimated in the semiclassical approach to be 
\be{nuclel}
\frac{\Gamma}{{\rm Volume}}\sim {\rm exp}\left(-\frac{E_{B}}{T}\right)\,,
\ee
where $E_{B}$ stands for the surface of the {\it baby strings}, which scales with the length as $E_{B}=T_{S}l$, $T$ being the energy scale of the baby string. On the other hand, within the context of our conjecture, the scalar constraint must describe creations and annihilations of SNS1-branes in non-perturbative regime beyond the semiclassical or perturbative string theory approximation.

\section{Conclusions and outlooks}
\noindent 
We conjectured the existence of a H-duality, which may unify topological M-theory and Loop Quantum Gravity, argueing that non-trivial gravitational holonomies can be put in correspondence with space-like M-branes. We grounded our proposal on the low wave-length limit of topological M-theory, showing how this latter re-constructs the theory of 3+1D gravity in the self-dual variables' formulation.\\

In our considerations, we have mainly discussed the H-duality between space-like NS1 foam and spinfoam, focusing more on the consequences of canonical quantization's techniques. Nonetheless it is still rather unclear whether our arguments can be more generically extended to ordinary M-branes, D-branes and NS-branes. In principle, it sounds reasonable to extend the duality and account for covariant quantization techniques. \\

A crucial problem is to understand how S-branes and D-branes fit in this picture. From the perturbative string theory point of view, S-branes and D-brane undergo a long-range interaction, exchanging closed strings. In the non-perturbative regime, this should correspond to an exchange of dilatons, gravitons and B-forms's excitations, entailing an infinite number of loops corrections. Furthermore, we should take into account also processes of brane instabilities, back-reacting on the system. We conjecture that these interactions are already encoded in the full non-perturbative regime realized on the LQG side. Nonetheless, we are urged to consider also extra graphs with respect to the flowers we focused on in this work. These and other features deserve a much deeper analysis, which we leave to forthcoming works.

\vspace{2cm} 

{\large \bf Acknowledgments} \\
\noindent
We are indebted to J. Lewandowski and H. Verlinde for enlightening discussions during Loops'17 in Warsaw, Poland. 



\vspace{1cm} 

\begin{thebibliography}{99}


\bibitem{BilsonThompson:2006yc} 
  S.~O.~Bilson-Thompson, F.~Markopoulou and L.~Smolin,
  Class.\ Quant.\ Grav.\  {\bf 24}, 3975 (2007)
  doi:10.1088/0264-9381/24/16/002
  [hep-th/0603022].
  
\bibitem{Alexander:2011jf} 
  S.~Alexander, A.~Marcian\`o and R.~A.~Tacchi,
  Phys.\ Lett.\ B {\bf 716}, 330 (2012)
  doi:10.1016/j.physletb.2012.07.034
  [arXiv:1105.3480 [gr-qc]].
  
\bibitem{Alexander:2012ge} 
  S.~Alexander, A.~Marciano and L.~Smolin,
  Phys.\ Rev.\ D {\bf 89}, no. 6, 065017 (2014)
  doi:10.1103/PhysRevD.89.065017
  [arXiv:1212.5246 [hep-th]].



\bibitem{Turiaci:2017zwd}
  G.~Turiaci and H.~Verlinde,
  arXiv:1701.00528 [hep-th].

\bibitem{Mertens:2017mtv}
  T.~G.~Mertens, G.~J.~Turiaci and H.~L.~Verlinde,
  arXiv:1705.08408 [hep-th].
  
\bibitem{Bianchi:2009tj}
  E.~Bianchi,
  Gen.\ Rel.\ Grav.\  {\bf 46} (2014) 1668
  doi:10.1007/s10714-014-1668-4
  [arXiv:0907.4388 [gr-qc]].
 
  
\bibitem{Dijkgraaf:2004te} 
  R.~Dijkgraaf, S.~Gukov, A.~Neitzke and C.~Vafa,
  Adv.\ Theor.\ Math.\ Phys.\  {\bf 9}, no. 4, 603 (2005)
  doi:10.4310/ATMP.2005.v9.n4.a5
  [hep-th/0411073].
  
\bibitem{Hitchin:2000jd}
  N.~J.~Hitchin,
  J.\ Diff.\ Geom.\  {\bf 55} (2000) no.3,  547
  [math/0010054 [math.DG]].
  
\bibitem{Hitchin:2001rw}
  N.~J.~Hitchin,
  math/0107101 [math-dg].
   

\bibitem{G2} E.~Bonan, 
C. R. Acad. Sci. Paris 262 (1966) 127;

\bibitem{G2a}
R.~L.~Bryant, 
{\it ``Metrics with exceptional holonomy''}, 
Annals of Mathematics (Annals of Mathematics) 126 (2) (1987) 525. 

\bibitem{G2b}
R.~L.~Bryant and S.~M.~Salamon, 
{\it ``On the construction of some complete metrics with exceptional holonomy''}, 
Duke Mathematical Journal 58 (1989) 829.

\bibitem{G2c}
M.~Fernandez and A.~Gray, 
{\it ``Riemannian manifolds with structure group G2''}, 
Ann. Mat. Pura Appl. 32 (1982) 19.

\bibitem{Joyce}  D.~.D.~Joyce,  
{\it``Compact Manifolds with Special Holonomy''}, 
Oxford Mathematical Monographs, Oxford University Press (2000).
  
\bibitem{HL}
R.~Harvey and H.~B.~Lawson, 
Acta Mathematica 148 (1982), 47.

\bibitem{Gukov}

S.~Gukov, 
{\it ``M-theory on manifolds with exceptional holonomy''}, Fortschr. Phys. 51 (2003), 719.


\bibitem{Smolin:2005gu}
  L.~Smolin,
  Nucl.\ Phys.\ B {\bf 739} (2006) 169
  doi:10.1016/j.nuclphysb.2006.01.016
  [hep-th/0503140].
  
\bibitem{Krasnov:2016wvc} 
  K.~Krasnov,
  Phys.\ Lett.\ B {\bf 772}, 300 (2017)
  doi:10.1016/j.physletb.2017.06.025
  [arXiv:1611.07849 [hep-th]].
  
\bibitem{Ling:2000dk}
  Y.~Ling and L.~Smolin,
  Nucl.\ Phys.\ B {\bf 601} (2001) 191
  doi:10.1016/S0550-3213(01)00063-3
  [hep-th/0003285].



\bibitem{Sen:2002nu}
  A.~Sen,
  JHEP {\bf 0204} (2002) 048
  doi:10.1088/1126-6708/2002/04/048
  [hep-th/0203211].
  
\bibitem{Sen:2002in}
  A.~Sen,
  JHEP {\bf 0207} (2002) 065
  doi:10.1088/1126-6708/2002/07/065
  [hep-th/0203265].


\bibitem{Gutperle:2002ai}
  M.~Gutperle and A.~Strominger,
  JHEP {\bf 0204} (2002) 018
  doi:10.1088/1126-6708/2002/04/018
  [hep-th/0202210].

\bibitem{Strominger:2002pc}
  A.~Strominger,
  Conf.\ Proc.\ C {\bf 0208124} (2002) 20
  [hep-th/0209090].
  
\bibitem{Hashimoto:2002sk}
  K.~Hashimoto, P.~M.~Ho and J.~E.~Wang,
  Phys.\ Rev.\ Lett.\  {\bf 90} (2003) 141601
  doi:10.1103/PhysRevLett.90.141601
  [hep-th/0211090].
  
\bibitem{Ohta:2003uw}
  N.~Ohta,
  Phys.\ Lett.\ B {\bf 558} (2003) 213
  doi:10.1016/S0370-2693(03)00274-0
  [hep-th/0301095].

\bibitem{Strominger:2001pn}
  A.~Strominger,
  JHEP {\bf 0110} (2001) 034
  doi:10.1088/1126-6708/2001/10/034
  [hep-th/0106113].




\bibitem{13}
H. Lu, C. Pope and K. Stelle, 
Nucl. Phys. {\bf B481} (1996) 313.





\bibitem{S11}
E. Cremmer, B. Julia and J. Scherk, 
Phys.Lett. {\bf B76} (1978) 409.

\bibitem{S11b}
M.J. Duff and K.S. Stelle, 
Phys.Lett. {\bf B253} (1991) 113.

\bibitem{L1}
E. Braaten, T. Curtright, G. Ghandour and C. B. Thorn, 
Phys. Rev. Lett. {\bf 51}, 19
(1983); Annals Phys. 153, 147 (1984).

\bibitem{L2}
J. Polchinski, 
UTTG-19-90 Presented at Strings '90 Conf., College Station, TX, Mar 12-17, 1990.


  
\bibitem{Livine:2013tsa}
  E.~R.~Livine,
  J.\ Math.\ Phys.\  {\bf 54} (2013) 123504
  doi:10.1063/1.4840635
  [arXiv:1307.2719 [math-ph]].
  
  
\bibitem{Livine:2013gna} 
  E.~R.~Livine,
  Class.\ Quant.\ Grav.\  {\bf 31}, 075004 (2014)
  doi:10.1088/0264-9381/31/7/075004
  [arXiv:1310.3362 [gr-qc]].
  


\bibitem{Charles:2016xwc}
  C.~Charles and E.~R.~Livine,
  Gen.\ Rel.\ Grav.\  {\bf 48} (2016) no.8,  113
  doi:10.1007/s10714-016-2107-5
  [arXiv:1603.01117 [gr-qc]].


\end{thebibliography}
\end{document}